\documentclass{aa}  

\usepackage{graphicx}
\usepackage{txfonts}
\usepackage{float}
\usepackage{makecell}
\usepackage{adjustbox}
\usepackage{rotating}
\usepackage[dvipsnames]{xcolor}
\usepackage{ulem}

\begin{document}

   \title{Tracing Stellar Populations through the Mg–Al Anti-Correlation in Gaia-ESO Globular Clusters}


   \author{H. Sinclair-Wentworth\inst{1}
          \and
          C.~C. Worley\inst{1}
          \and
          P. Jofr\'{e} \inst{2}
          \and
          J. Schiappacasse-Ulloa\inst{3} 
          \and
          L. Magrini\inst{3}
          }


   \institute{School of Physical and Chemical Sciences - Te Kura Mat\={u}, University of Canterbury, New Zealand\\
              \email{heather.sinclair-wentworth@pg.canterbury.ac.nz}
   \and
   Instituto de Estudios Astrof\'{i}sicos, Facultad de Ingenier\'{i}a y Ciencias, Universidad Diego Portales, Ej\'{e}rcito Libertador 441, Santiago, Chile
   \and 
   INAF–Osservatorio Astrofisico di Arcetri, Largo Enrico Fermi 5, 50125 Florence, Italy.
   }

   \date{Received ,2026; accepted ,2026}

  \abstract
  {Globular clusters (GCs) are fragments of the chemical history and evolution of the Milky Way. Their notable intra-cluster abundance variations have been used to characterise chemical trends and stellar populations. However, the origins of these populations and GCs in general are still largely unknown.}
  {Our aim is to explore the origin of the stellar populations belonging to 14 GCs making use of their abundances derived from the Gaia-ESO Survey (GES) spectra.
  In particular, we use the Mg-Al anti-correlation to separate the populations.}
   {The GES spectra were revisited to derive additional homogenous chemical abundances for Mg and Al. A Gaussian Mixture Model (GMM) was applied to the [Mg/Al] distribution for each cluster to define a threshold between the two main populations.}
  {Eight clusters imply the presence of Mg-Al anti-correlations and dual populations. We find there is a regime transition near [Fe/H]$\sim -$1.4, where, the Mg-Al anti-correlation disappears at higher metallicity. Several clusters, regardless of their Mg-Al anti-correlation behaviour, exhibit metallicity spreads (e.g., NGC~2808, NGC~6752) and $s$-process variations (e.g., NGC~7078, NGC~1851), suggesting complex enrichment histories.}
  {The presence and spread of the Mg-Al anti-correlation depend on cluster mass and metallicity, with metal poor and/or massive clusters exhibiting the clearest anti-correlation. This work demonstrates additional abundances that can be retrieved from Gaia-ESO Survey spectra and provides further insight into the chemistry of multiple stellar populations within GCs. These findings highlight the chemical complexity of GCs and the need for large, homogeneous datasets to trace their formation histories.}

   \maketitle

\section{Introduction}

Globular clusters (GCs) are ancient, dense clusters of stars that contain key information about the chemical history of the Milky Way. These cosmic fossils were originally thought to be simple stellar populations, comprised of stars with similar ages and abundances. However, more complex chemical trends have been observed with large intra-cluster variations and anti-correlations (e.g Na-O, Mg-Al), indicating GCs can have multiple stellar populations (MSP) \citep{abund_vari, mult_stell_pop}. 

A prominent chemical feature of most GCs is the presence of anti-correlations, in particular, the sodium-oxygen (Na-O) and magnesium-aluminium (Mg-Al) anti-correlations \citep{gratton_anti_corr, NaOVII, Pan_mgal}. Both the Na-O and Mg-Al anti-correlations appear to be related to the metallicity and mass of a GC. Specifically, the metal poor and massive clusters show a larger spread in the Mg-Al distribution \citep{Pan_mgal}, while massive clusters exhibit more extreme O depletion in the Na-O anti-correlation \citep{NaOII}.

The cause of these observed chemical characteristics remains inconclusive with contribution from multiple astrophysical sources considered possible. High-temperature H-burning during the CNO, Ne-Na, and Mg-Al cycles is believed to be the likely origin of the observed light element variations. This H-burning could occur in intermediate-mass Asymptotic Giant Branch (AGB) stars, Fast Rotating Massive Stars (FRMS), or Very Massive Stars (VMS) \citep{deep_mix1993, mult_stell_pop, MP_clust}. These stars could have existed as the first generation of stars which then pollute the intra-cluster material after their deaths, providing enriched material from which the second generation of stars are born. Each of these pollution scenarios require at least one round of dilution with pristine material to reproduce observed abundances \citep{dercole2016, mult_stell_pop}. No single model explains both the Na-O and Mg-Al anti-correlations simultaneously, leaving key abundance patterns unresolved. 

Three primary stellar populations are commonly observed in GCs, defined by \citet{NaOVII} through the Na-O anti-correlation. These are the Primordial (\textit{P}), Intermediate (\textit{I}) and Extreme (\textit{E}) populations, where the \textit{I} and \textit{E} populations show enrichment in Na and depletion in O compared to the \textit{P} population \citep{star_gen_GC}. The \textit{P} and \textit{I} populations are present in all GCs with MSPs, with the \textit{I} population comprising the majority of stars ($50-70\%$) within a GC, and the \textit{P} population making up about $30\%$. The \textit{E} population however, is not always present \citep{star_gen_GC}.

Many spectroscopic surveys have observed GCs, including the Gaia-ESO Survey  \citep[GES,][]{GES_Motivation, randich2022GES}, GALactic Archaeology with HERMES \citep[GALAH,][]{GALAH}, and the Apache Point Observatory Galactic Evolution Experiment \citep[APOGEE,][]{APOGEE}. The analysis presented here uses the GC sample from the final release of the Gaia-ESO Survey, available in the ESO archive\footnote{\url{https://www.eso.org/qi/catalogQuery/index/411}}.

GES is a large public spectroscopic survey of approximately 100,000 stars from multiple regions in the Milky Way such as the halo, disk and bulge, including also a variety of open and globular clusters,  that cover a wide variety of stellar parameters and spectral types \citep{GAIA_ESO}. The GES observations utilise the Fibre Large Array Multi Element Spectrograph (FLAMES) on European Southern Observatory's (ESO) Very Large Telescope (VLT). FLAMES includes both the medium resolution ($R$) spectrograph GIRAFFE ($R \simeq 20 000$) \citep{FLAMES}, and high resolution spectrograph, the Ultraviolet and Visual Echelle Spectrograph (UVES) ($R \simeq 47 000$) \citep{UVES}. A homogeneous catalogue of astrophysical parameters, chemical abundances and radial velocities has been created from this survey to further our understanding of galactic and stellar evolution \citep{GES_Motivation, GES_hom_23}. 

Within GES there is a sample of 3998 stars, not including any membership cut, observed across 14 GCs fields which have already been the subject of several studies. In particular, the exploration of the lighter elements has been undertaken for GES DR4 in \citet{Pan_mgal}, where both the Mg-Al and Na-O anti-correlations were examined for 9 GCs. Albeit not with GES data, \cite{jose2024} made an initial study with UVES spectra to characterise GCs in terms of n-capture elements. This work was extended in \cite{jose2025} to use GES DR5 to show relationships between these heavier elements and the lighter elements typically used to define MPs. There have been individual studies of NGC~4372 \citep{ges_4372} and NGC~1851 \citep{ges_1851}, each which investigated chemical abundance distributions of the cluster members using GES data.

While the GES catalogue contains an extensive collection of chemical abundances, the set for the GC sample is sparse and incomplete. Thus, for this study, the GES GC spectra were revisited to expand the set of measured abundances for Mg and Al. The analysis was carried out ensuring homogeneity with the GES DR5 GC dataset. This has allowed us to explore in more depth the Mg-Al anti-correlation and its relationship to the two prominent populations within 14 GCs, specifically, the \textit{P} and \textit{I} populations. Spread in metallicity and heavier elements have been observed in some clusters, thus we also examined the GES [Fe/H] and the Y and Ba abundances as provided in \citet{jose2025} as potential evidence of multiple populations.

The structure of this paper is as follows: Section \ref{sec:data_member} outlines the curation of the GES sample and stellar membership selection; Section \ref{sec:params} outlines the GES parameters and corrections used for this sample ; Section \ref{sec:derive} and \ref{sec:mgal_ac} explains the derivation process and analysis of abundances, including the Mg-Al anti-correlation; Section \ref{sec:stat_method} presents the statistical methods used in this work; Section \ref{sec:dual_pops} shows the results of these statistical methods applied to the Mg-Al anti-correlation with two populations defined and their metallicities explored; Section \ref{sec:global_char} covers the global cluster characteristics such as metallicity and $s-$process element spread; Section \ref{sec:discussion} discusses our work in the context of literature and Section \ref{sec:conc} summarises this work.

\section{Data Curation and Sample Selection}
\label{sec:data_member}

The GES GCs cover a broad range of metallicities from $-2.37$ to $-0.49$ dex. The selection criteria and observing strategy for these clusters is described in \citet{Pan_cali}. In this work, we draw on the GES Data Release 5 (DR5; \citealt{GES_hom_23}) to select GC member stars for which we will re-derive and expand the Mg and Al abundance dataset. 

The GES GC sample contains measurements from GIRAFFE and UVES spectra, therefore each individual GC star could have spectra from either or both instruments. If a star was observed with both instruments, measurements from UVES spectra were given higher priority for inclusion in the final GES catalogue \citep{GES_hom_23}. For this work, the UVES spectra were also then selected as the preferred spectra for abundance derivation. 

GIRAFFE spectra from the ESO archive were also included in GES. However, as they were often not observed with sufficient quality criteria to match the GES conditions, the derived measurements had the lowest priority for selection. The analysis of these GC stellar spectra as part of the large dataset of F, G and K type stars for GES DR5 is described in \citet{Worley_2024}. Stars with poor-quality spectra were removed from the sample. Specifically, we excluded stars with an average S/N$<$30 across all available spectra and individual spectra with S/N$<$40.

To avoid contamination from Milky Way halo field stars, membership probabilities for individual stars were adopted from \citet{DR3_members}. Stars with membership probabilities greater than or equal to $0.8$ were considered cluster members, while those with lower probabilities were discarded. This selection criterion ensures that the analysed sample is highly likely to represent the chemical properties of the cluster population.

The majority of stars lie along the Red Giant Branch (RGB) of each cluster with some of the clusters showing a few stars on the Asymptotic Giant Branch (AGB) \cite[See figure 13 in][]{GES_hom_23}. NGC~5927 is the exception to this with a large sample of $\sim$ 50 Horizontal Branch (HB) stars observed out of the total $\sim$70. All stellar parameters and abundances were obtained consistently in GES and in this study.

\section{Astrophysical Parameters}
\label{sec:params}

Stars without astrophysical parameters in GES such as effective temperature ($T_{\text{eff}}$), surface gravity ($\log g$) and metallicity ([Fe/H]) were also excluded as these parameters are required for determining chemical abundances. The homogenisation of stellar parameters and use of the GC sample within the wider survey is described in \citet{GES_hom_23}. 

Microturbulence (v$_{\text{t}}$) values were not provided by \citet{GES_hom_23} for GIRAFFE stars. For this work, microturbulence values were calculated using relations derived from the GES UVES sample as described in \citet{smiljanic2014_ges}. Microturbulence values for the GC stars ranged from 0.9 - 1.95 km s$^{-1}$.

Outliers were removed based on the following criteria: Stars that lay outside the predominant parameter range for each cluster (typically 4400$<T_{\text{eff}}< $5200 and $\log g< $3); stars with [Fe/H] errors $>$ 0.1 from UVES if there were more than 5 UVES stars would remain; Stars with an absolute difference to the median [Fe/H] greater than 0.5 dex.

The $T_{\text{eff}}$ and $\log g$ values provided by GES were adopted without modification. However, trends in metallicity were found with both $T_{\text{eff}}$ and $\log g$ for several clusters. The corrections applied to their metallicity are described in Section \ref{apdx:feh_corr}.

\section{Abundance Derivation and Analysis}
\label{sec:derive}

In this work we have expanded the abundance dataset for the GES GCs by including additional Mg and Al abundances. These abundances were derived using synthetic spectral fitting with iSpec \citep{DetermineIspec, Speccaveats}, and the spectral synthesis code MOOG \citep{MOOG}. To remain consistent with GES, the line list from \citet{GESlinelist2020}, MARCS model atmosphere \citep{MARCS} and solar abundances from \citet{grev_solar} were used. 

\subsection{Element Selection}

The Na-O anti-correlation has historically been the primary chemical characteristic used for defining populations in GCs \citep{NaOVII}. Due to the GIRAFFE setup selection for GES, there is an insufficient number of GC stars with both Na and O measurements in DR5. Na and O lines are only present in UVES spectra, narrowing the sample significantly and within that subset, not all stars have reliable Na and/or O abundances due to low signal-to-noise. 
In contrast, Mg and Al spectral features are present in both the UVES and GIRAFFE spectra, allowing abundance measurements for a much larger sample of stars. The Mg-Al anti-correlation therefore provides an alternative diagnostic for defining stellar populations in GCs and forms the basis of this study. Accordingly, Mg and Al abundances were derived from both the UVES and GIRAFFE GES spectra.

\subsection{Atomic Line List}

The \citet{GESlinelist2020} line list includes flags for line blending properties and quality of transition probability, $\log gf$, noted as \textit{synflag} and \textit{gfflag} respectively. The labels are ``Y'' for yes,``U'' for uncertain, and ``N'' for no when assessing the usage of the line. Only lines with \textit{synthflag} and \textit{gfflag} as either `U' or `Y' were considered in order to avoid inaccurate and low quality line measurements. 

 Three individual Mg lines at $\sim$631 nm frequently produced overestimated abundances and were therefore excluded. From the full list of available lines in \citet{GESlinelist2020}, five Al~\textsc{i} and eight Mg~\textsc{i} were used in this study. Table ~\ref{tab:linelist} presents the final selection of lines and their corresponding wavelength.

\begin{table}[H]
    \centering
    \begin{tabular}{|c|p{6cm}|}
        \hline
        Element & Wavelength (nm) \\ \hline
        Mg \textsc{i} & 516.73, 517.27, 518.36, 552.84, 571.11,  871.78, 873.60, 880.68 \\ \hline
        Al \textsc{i} & 555.71, 669.60, 669.87, 877.29, 877.39 \\ \hline
    \end{tabular}
    \caption{Mg and Al spectral lines used for abundance derivation}
    \label{tab:linelist}
\end{table}

\subsection{Error Calculations}

iSpec provides an error for each individual line abundance measurement from the spectral synthesis fit, where an extremely small or large error indicates an inaccurate line fit. Any line abundance with an error $\le$ 0 dex or $\ge$ 3 dex in absolute abundance was excluded.

The errors that are reported with the final abundances for each star are the standard error of the individual line abundances. This was calculated as the standard deviation ($\sigma$) divided by the square root of the number of lines, and multiplied by $\sqrt{\pi/2}$ to obtain the standard error of the median, since the final abundances are the median of all individual lines. Uncertainties for elements with only one line were taken as the error of the line fit from iSpec.

\subsection{Scaling}
\label{sec:scaling}

Within GES, medium resolution results were scaled to the high resolution ones to form a homogeneous set on a common reference frame, as described in \cite{Worley_2024}. Therefore, the individual line abundance values measured with iSpec were calibrated to the GES abundances scale to ensure consistency between our work and GES. For every star in a cluster, the difference between the abundance from each individual line and that star’s global GES abundance was calculated. This produced a set of difference values for each spectral line across all stars in the cluster. The median of these values was then taken as the line-specific offset and applied to each star's corresponding line abundance to place it on the GES scale.

In cases where only a small number of GES abundances were available, the GES reference frame was less well defined for adjusting iSpec abundances. This occurred for NGC~4590, where only one star existed in GES with an Al abundance. In this case, the difference between the iSpec individual line values and the single GES abundance was still calculated and applied. A comparison with clusters of similar metallicity (NGC~4372 and NGC~7078) showed that the resulting Al distribution was consistent, and therefore the single star scaling provided an adequate approximation.

\subsection{Abundance Outlier Removal}

The bulk of the abundance distributions across all clusters was found to lie below 1.2 dex for [Mg~\textsc{i}/Fe] and below 1.9 dex for [Al~\textsc{i}/Fe]. Abundances above these limits were flagged as potential outliers and excluded from further analysis, as they are likely affected by unreliable spectral line fitting, especially in metal-poor stars where absorption features are weak.

\subsection{Final Abundance Dataset}
\label{sec:final_abunds}

The final abundance dataset of 1135 individual stars across 14 GCs was used to explore abundance trends. Table \ref{tab:clust_stars} shows the number of stars per cluster with original GES Mg and Al abundances (GES N$_{*}$), number of stars with Mg and Al measurements from this work (N$_{*}$), cluster metallicity from \citet{harris}, and mass from \citet{mass_cat}. The majority of the stars have medium-resolution GIRAFFE spectra (907 stars), while high-resolution spectra are available for 228 stars. The typical signal to noise (S/N) for the stars included in this dataset is $\sim$70 for both UVES and GIRAFFE. NGC~5927 is the only GC for which there are fewer stars in this study than the original GES GC dataset. 4 stars were excluded due to S/N $<$30. The table of final abundances can be found in Table \ref{tab:example_dataset} and is available in full online.

\renewcommand{\arraystretch}{1.1} 

\begin{table}
\centering
\caption{Summary of cluster properties, including the number of stars with Mg and Al abundances from GES (GES N$_{*}$) and from this study (This Study $N_{*}$), the metallicity ([Fe/H]) \citep{harris} and the present day cluster mass ($M_{\odot}$) \citep{mass_cat}. Clusters are ordered by metallicity.}
\begin{tabular}{|l|c|c|c|c|}
\hline
Cluster & GES $N_{*}$ & This Study N$_{*}$ & [Fe/H] & Mass ($M_{\odot}$) \\ \hline \hline

NGC7078 & 5 & 82 & -2.40 & $4.53\times10^{5}$ \\\hline
NGC4590 & 1 & 68 & -2.37 & $1.23\times10^{5}$ \\\hline
NGC4372 & 9 & 86 & -2.20 & $2.49\times10^{5}$ \\\hline
NGC4833 & 14 & 43 & -1.97 & $2.47\times10^{5}$ \\\hline
NGC1904 & 49 & 70 & -1.57 & $1.69\times10^{5}$ \\\hline
NGC6752 & 79 & 116 & -1.55 & $2.39\times10^{5}$\\\hline
NGC7089 & 71 & 81 & -1.51 & $5.82\times10^{5}$\\\hline
NGC6218 & 80 & 81 & -1.25 & $8.65\times10^{4}$ \\\hline
NGC1261 & 33 & 43 & -1.16 & $1.67\times10^{5}$ \\\hline
 NGC362 & 94 & 97 & -1.08 & $3.45\times10^{5}$ \\\hline
NGC1851 & 80 & 90 & -1.07 & $3.02\times10^{5}$ \\\hline
NGC2808 & 62 & 63 & -1.07 & $7.42\times10^{5}$ \\\hline
 NGC104 & 142 & 142 & -0.76 & $7.79\times10^{5}$ \\\hline
NGC5927 & 77 & 73 & -0.43 & $3.54\times10^{5}$ \\\hline

\end{tabular}
\label{tab:clust_stars}
\end{table}

\section{The Mg-Al Anti-Correlation}
\label{sec:mgal_ac}
 
While an anti-correlation generally refers to a trend where one abundance increases as the other decreases, as observed for Na-O, the Mg-Al anti-correlation is often characterised by a spread in Mg and Al abundances, with variations occurring in either or both elements. Typically, there is also a population of stars with [Al/Fe]$>0.6$ present as seen in \citet{carretta_2014} for NGC~2808. This enhanced Al population appears to have a large spread in Mg.

\begin{figure}
    \centering
    \includegraphics[width=0.75\linewidth]{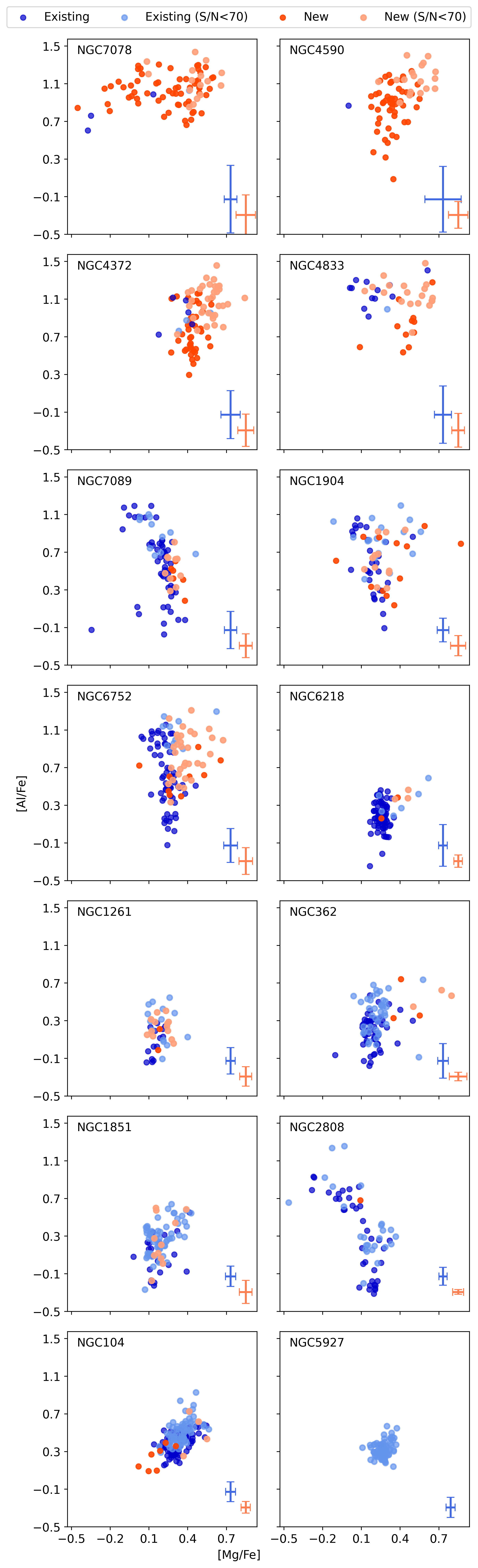}
    \caption{Mg-Al anti-correlation for all GES clusters, ordered from left to right, top to bottom in increasing metallicity. Stars with existing GES measurements are in blue while those with new iSpec measurements are in orange. The lighter shades of orange and blue indicate stars with S/N<70.}
    \label{fig:mgal_anticor}
\end{figure}

Figure \ref{fig:mgal_anticor} shows the Mg-Al distribution for each cluster from the abundances derived in this work. Stars that already had Mg and Al measurements available in the GES sample are shown in blue, while stars for which Mg and Al abundances are newly measured in this work are shown in orange. In total, 353 stars were added in this analysis with Mg and Al values that were previously unavailable. 

The stars without GES Mg and Al measurements generally exhibit larger abundance uncertainties than those with GES existing measurements. This is likely due to the fact that for the newly analysed stars the abundance determination is predominantly done with medium resolution GIRAFFE spectra, and the median signal-to-noise ratio for these stars is typically lower than that of the stars with available GES abundances. Although the orange points in Figure \ref{fig:mgal_anticor} exhibit a slightly increased spread compared to the blue points, the derived Mg and Al abundances are in very good agreement between stars with and without GES measurements across the majority of clusters. 

The distribution of Mg and Al is seen to change as metallicity increases, with the more metal poor clusters exhibiting a wider distribution for both elements than the metal rich, which appear very tight. This is expected and aligns well with previous works such as \citet{NaOVIII, Pan_mgal, meszaros2020}, where a trend of decreasing Mg-Al spread with metallicity is observed. Most of the metal rich clusters, NGC~6218, NGC~1261, NGC~362, NGC~1851, NGC~104 and NGC~5927, all have a relatively small spread in Mg and Al, going to tighter distributions as metallicity increases. NGC~2808 is an outlier in this metallicity trend as it displays a strong anti-correlation with two main components. The possible cause of this is discussed in Section~\ref{sec:mgal_spread}. 

Metal poor clusters NGC~7078, NGC~4590, NGC~4372 and NGC~4833 all show a spread in Mg and Al (where Al is notably enhanced), whereas NGC~7089, NGC~1904, NGC~6752 and NGC~2808 display the more typical anti-correlation distribution with the greater Al spread. To evaluate the presence of a statistically robust anti-correlation, a p-value and Pearson Correlation Coefficient (PCC) were computed for each cluster. The anti-correlations in NGC~7089 and NGC~2808 both have a p-value less than 0.01 and negative PCC, indicating these are the only two clusters with a statistically detectable anti-correlation.

These anti-correlations have previously been noted in \citet{Pan_mgal, meszaros2020} for NGC~7089 and many studies for NGC~2808 including \citet{NaOVIII, Pan_mgal, meszaros2020}. After removing stars with S/N$<$70 from NGC~6752, a statistically significant anti-correlation is recovered, in agreement with previous studies (e.g., \citet{starB_abundances, NaOVIII, Pan_mgal, meszaros2020}). While the other metal poor clusters do not show a statistically significant anti-correlation, their morphology is distinctly different to the metal rich clusters, i.e enhanced in Al, as seen in Figure\ref{fig:mgal_anticor}.

\begin{figure*}
    \centering
    \includegraphics[width=17cm]{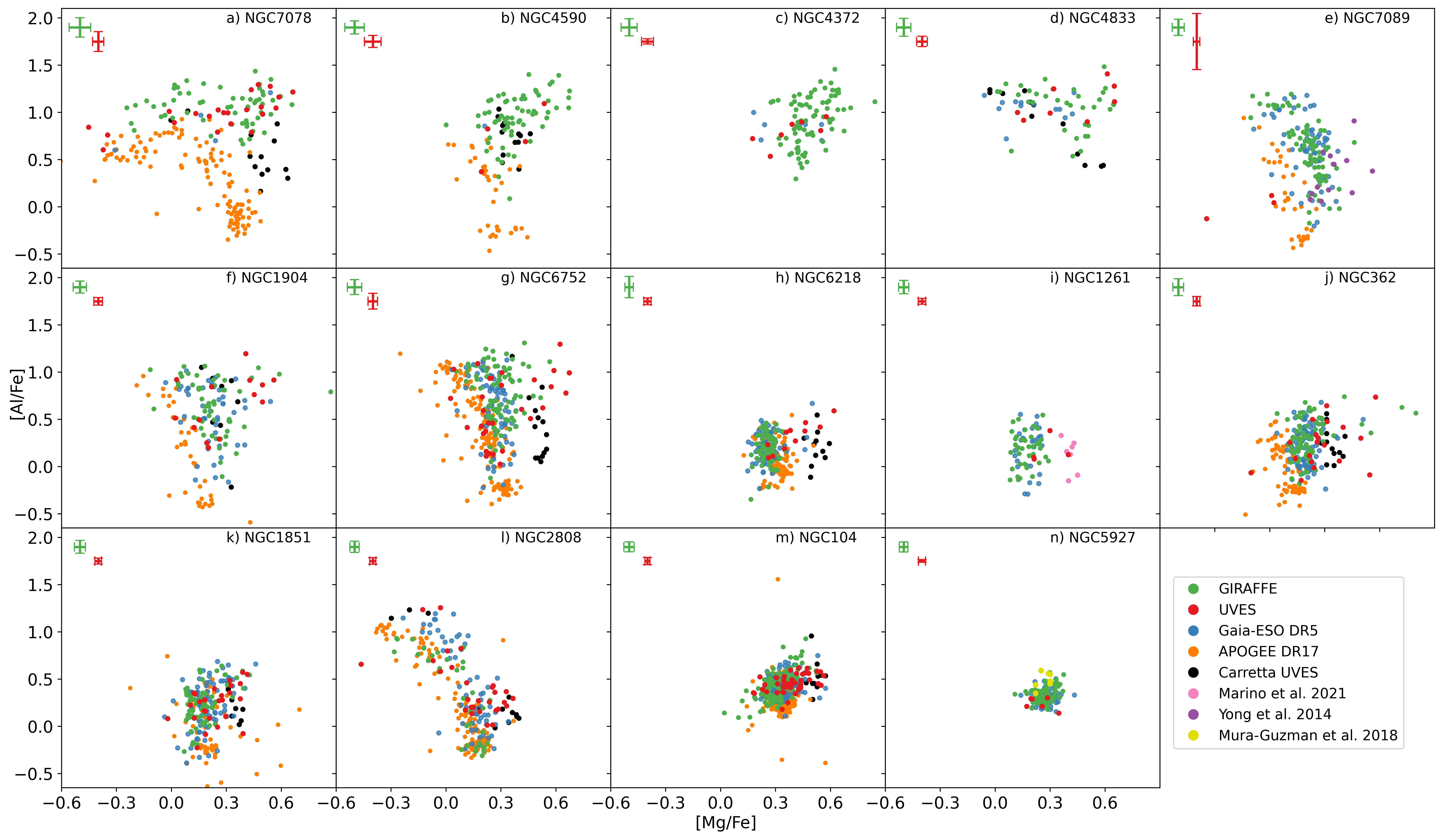}
    \caption{Mg-Al distributions for GIRAFFE (green) and UVES (red) from this work. Results from APOGEE DR17 \citep{schiavon2024apogee} are in orange and Gaia-ESO \citep{GES_hom_23} in blue. Values from the UVES GC Survey by Carretta et al are in black \citep{NaOVIII, carretta_1851, carretta2013_362, carretta_4833}. The individual studies by \citet{marino2021_1261} on NGC1261 (pink), \citet{yong2014_m2} on NGC7089 (purple) and \citet{muraguzman2018_5927} on NGC5927 are also shown. Clusters are in increasing metallicity order. Typical errors for both GIRAFFE and UVES from this work are in the top left corners.}
    \label{fig:mgal_with_literature}
\end{figure*}

A direct comparison of this work with the literature for the Mg-Al distributions for each cluster can be seen in Figure \ref{fig:mgal_with_literature}. APOGEE DR17 \citep{schiavon2024apogee}, GES DR5 \citep{GES_hom_23} and Carretta et al's FLAMES GC survey \citep{NaOVIII, carretta_1851, carretta2013_362, carretta_4833} provide values for most of the GCs in this work. Additionally, single cluster papers such as \citet{marino2021_1261, yong2014_m2, muraguzman2018_5927} are included for three clusters not part of those larger surveys. NGC~4372 does not appear to have any other Mg and Al literature values except for the GES study in \citet{ges_4372}. 

At the metal-poor end, an offset appears to be present between the optical studies and the infrared APOGEE abundance measurements. In particular, NGC~7078 and NGC~4590 (Figure \ref{fig:mgal_with_literature}a and b) exhibit the largest discrepancies in their abundance distributions when compared with values reported in the literature. The low [Al/Fe], high [Mg/Al] population found in both APOGEE and the FLAMES GC survey is not observed in this work. Instead, there is a clumping of stars with high [Mg/Fe] and high [Al/Fe] from both UVES and GIRAFFE. This includes the stars with S/N < 70 as mentioned for Figure \ref{fig:mgal_anticor}. In NGC~7078, the population of stars with [Mg/Fe]$<$0 identified in our dataset is also present in APOGEE, suggesting that this feature is real.

In NGC~7078 there are 42 stars in common between our sample and APOGEE. Directly comparing the parameters, there is a median difference of $-$90K in T$_{\text{eff}}$, $-$0.42 in $\log(\text{g})$ and $-$0.09 in [Fe/H]. Only 7 stars are common between the two samples in NGC~4590, which represents a small sample but exhibits consistent offsets in the stellar parameters ($-$170 K in T$_{\text{eff}}$, $-$0.68 dex in $\log(\text{g})$, and $-$0.07 dex in [Fe/H]). The offset, particularly in $\log(\text{g})$, is likely responsible for part of the observed abundance discrepancies.

\section{The [Mg/Al] Ratio as a Population Threshold}
\label{sec:stat_method}

Using the final sample of stars, defined in Section \ref{sec:final_abunds} with S/N$>$30, we see extent of the Mg and Al abundance spreads varies between clusters, and the Mg-Al anti-correlation is not universally detected. Based on this, the definition of stellar populations should consider both elements simultaneously. The [Mg/Al] ratio therefore offers an effective means of combining Mg and Al into a single parameter, enabling the identification of multiple abundance groups.  To do this, probability density plots are created from the [Mg/Al] ratio. Using a Gaussian Mixture Model (GMM), multiple components can be searched for in the [Mg/Al] density plots, corresponding to multiple populations. 

Previous analysis of GCs has detected multiple stellar populations within the majority of GCs, with exceptions seen in low mass ($\lesssim$10$^{5}$ M$_{\odot}$) or young (< 2 Gyr) clusters \citep{mult_stell_pop}. In this work, a minimum of two populations are assumed for each GC. Multiple populations are known to exist in some clusters; for example, NGC~2808 has been observed to host at least five distinct populations \citep{2808_5MP_2015, milone_2808, 2808_5MP_2016, 2808_5MP_2018}. Therefore, we consider models containing up to five populations, consistent with the highest number of populations currently identified in Galactic globular clusters.

Initially, we combined all the GC stars and applied GMM to find a global threshold. The GMM was run multiple times, with a different number of components (2 to 5). The Bayes Information Criterion (BIC) of each model was compared to find the lowest value, indicating the best fit. Two components were found to provide the best fit with the lowest BIC for the combined cluster sample.

\begin{figure}
    \centering
    \includegraphics[width=0.8\linewidth]{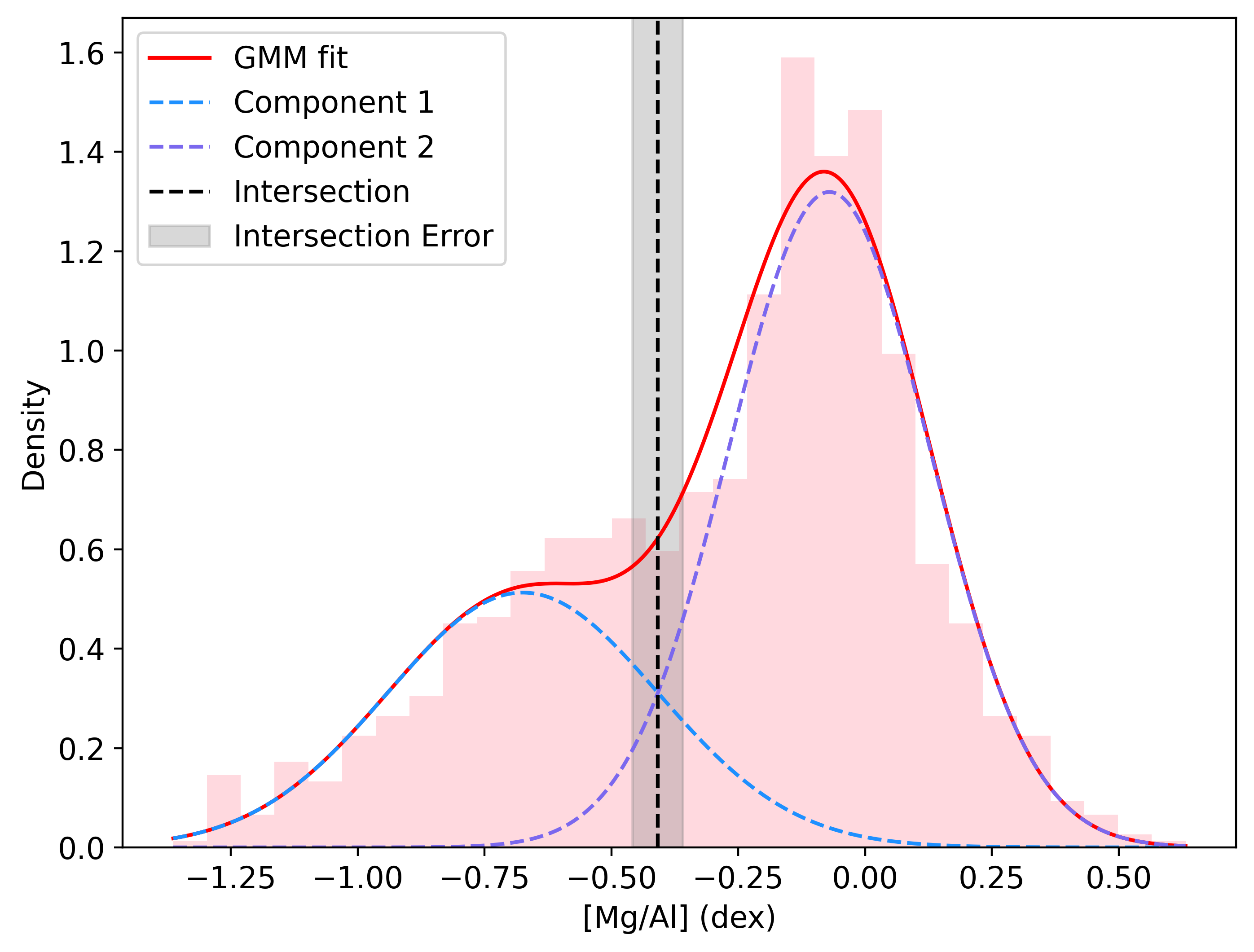}
    \caption{Probability density of [Mg/Al] abundances for all stars in pink. Two Gaussian components fitted to the data are indicated in blue and green, with the overall GMM fit in red. The black dashed line shows the intersection of the two components, with the grey shaded region indicating the uncertainty on this intersection.}
    \label{fig:mgal_density_whole}
\end{figure}

This can be seen in Figure \ref{fig:mgal_density_whole} where two main peaks at [Mg/Al]$\approx -0.75$ and $\approx -0.1$ correspond to the two populations. This global intersection was found to be at [Mg/Al] $=$ -0.41 $\pm$ 0.05. The intersection of these two components could therefore be used as a threshold to split the stars into two populations. The uncertainty on the intersection was estimated as the standard deviation of the range of intersection values found using a bootstrap method which ran the GMM 100 times. From this global approach, the first component contains 366 stars while the second one has 769.

While using the entire dataset provides a generalised separation between populations, it obscures the unique and potentially different range of values for each individual cluster. To provide a more tailored threshold between the populations, this analysis was repeated independently for each GC. As before, the number of optimal components for each cluster was decided using the BIC. For all clusters except NGC~7089, two components were found to provide the lowest BIC value.

For NGC~7089, the GMM indicated that a three-component model provided the optimal fit. However the intersection value for the lowest [Mg/Al] component was comparable to the single intersection value for clusters of similar metallicity when fitting two components. Therefore, for ease of comparison across the GCs, NGC~7089 will also be split into only two populations using the intersection value for the lowest [Mg/Al] component. Investigating the nature of the third possible component is left to future work.

\begin{figure}
    \centering
    \includegraphics[width=\linewidth]{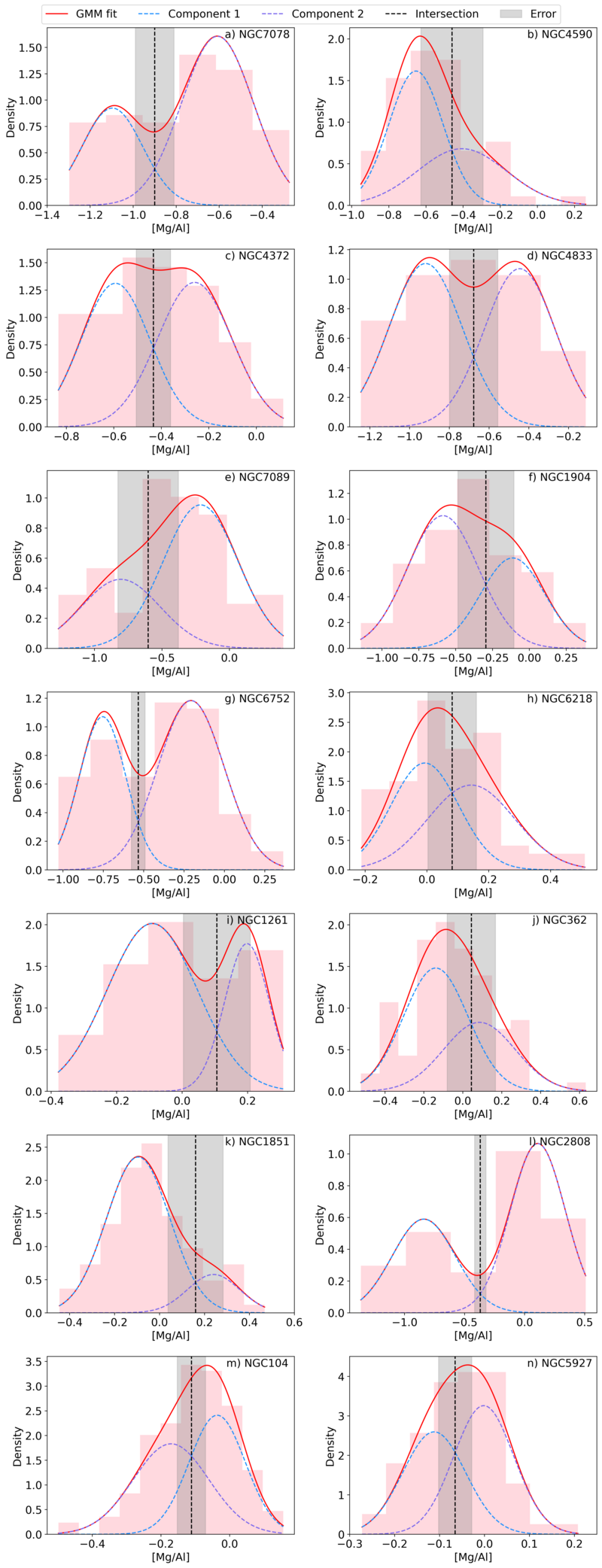}
    \caption{Same as Figure \ref{fig:mgal_density_whole} except for each individual cluster, ordered by increasing metallicity from left to right and top to bottom.}
    \label{fig:mgal_density}
\end{figure}

Figure~\ref{fig:mgal_density} displays the probability density and threshold value found for each cluster using two components for the GMM. The intersections and corresponding error for each cluster are shown in Table \ref{tab:mgal_intersections}. The large variation in the intersection points confirms that this method better captures the individual characteristics of each cluster.

The metal rich clusters, with the exception of NGC~2808, all exhibit a much smaller range of [Mg/Al] (See Figure~\ref{fig:mgal_density} h, i, k, n). While the GMM is able to separate these metal-rich clusters into two populations, their overall abundance distributions remain very compact, with the intersection point occurring at higher [Mg/Al] values than in the other clusters. The intersections also do not align well with the global estimate from Figure~\ref{fig:mgal_density_whole}. Therefore, the clusters NGC~6218, NGC~1261, NGC~362, NGC~1851, NGC~104 and NGC~5927 were deemed to not show sufficient variation in [Mg/Fe] and [Al/Fe] to detect separate populations as found in the more metal poor clusters.

While we do not detect a large Mg-Al variation or anti-correlation, previous studies have shown these clusters do show the Na-O anti-correlation. These large variations in Na and O imply that multiple populations exist in these cluster. This indicates both the Na-O and Mg-Al anti-correlations have distinct, different behaviours and is discussed later in Section \ref{sec:discussion}.

The remaining clusters, NGC~7078, NGC~4590, NGC~4372, NGC~4833, NGC~7089, NGC~1904, NGC~6752 and NGC~2808, all display two distinct populations separated by a threshold, and will hereafter be referred to as the dual-population clusters. Their threshold values, listed in Table \ref{tab:mgal_intersections}, vary between clusters but span 0.53 dex. This  spread indicates any additional GC showing a Mg-Al anti-correlation would likely exhibit a threshold in a similar range.

\begin{table}
\centering
\caption{Intersection and intersection error for each cluster using the [Mg/Al] population split}

\begin{tabular}{|l|c|c|}
\hline
Cluster & \makecell{[Mg/Al] \\ Intersection} & \makecell{[Mg/Al] \\ Intersection Error} \\ \hline \hline
NGC7078 & -0.90 & 0.09\\\hline
NGC4590 & -0.46 & 0.17\\\hline
NGC4372 & -0.43 & 0.07\\\hline
NGC4833 & -0.68 & 0.12\\\hline
NGC7089 & -0.60 & 0.22\\\hline
NGC1904 & -0.29 & 0.19\\\hline
NGC6752 & -0.53 & 0.04\\\hline
NGC6218 & 0.08 & 0.08\\\hline
NGC1261 & 0.11 & 0.10\\\hline
NGC362 & 0.04 & 0.12 \\\hline
NGC1851 & 0.16 & 0.12 \\\hline
NGC2808 & -0.37 & 0.05\\\hline
NGC104 & -0.11 & 0.04 \\\hline
NGC5927 & -0.07 & 0.04 \\\hline
\end{tabular}
\label{tab:mgal_intersections}
\end{table}

\section{Dual Population Clusters}
\label{sec:dual_pops}

\subsection{Classification of Populations: Primordial and Intermediate}

Using the [Mg/Al] ratio to represent the variation in Mg-Al and applying GMM, we have defined two stellar populations within NGC~7078, NGC~4590, NGC~4372, NGC~4833, NGC~7089, NGC~1904, NGC~6752 and NGC~2808. These two populations can be compared to those found from previous studies. Specifically, the Primordial (\textit{P}), Intermediate (\textit{I}) and Extreme (\textit{E}) populations defined in \citet{NaOVII}, hereafter C09b, using the Na-O anti-correlation for 17 GCs. 

\begin{figure*}
    \centering
    \includegraphics[width=17cm]{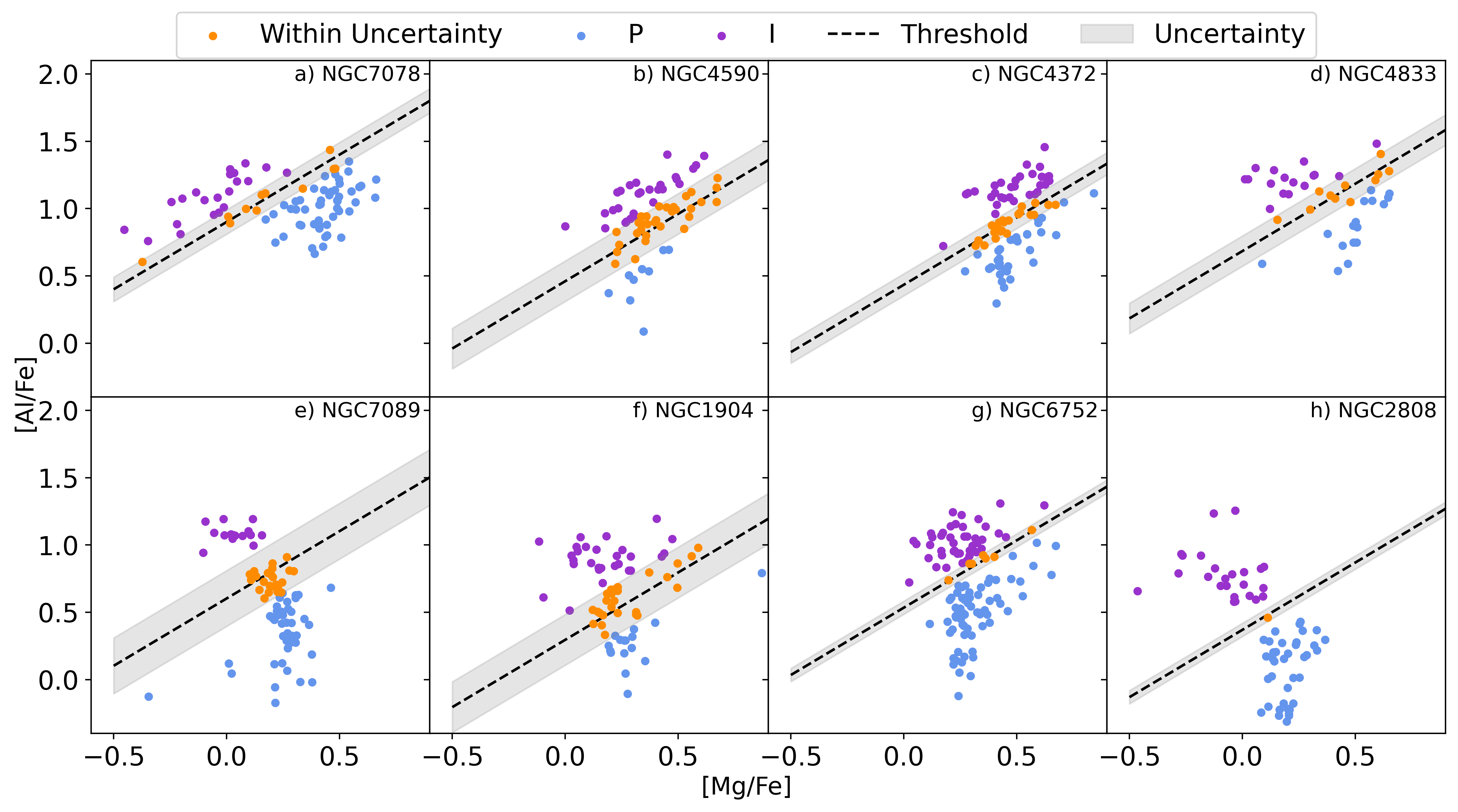}
    \caption{Mg-Al anti-correlation for the dual population clusters. Blue indicates the P population, purple the I population and orange the stars within the uncertainty on the threshold. The threshold itself is represented by the black dashed line with the uncertainty region in grey.}
    \label{fig:mgal_P_I}
\end{figure*}

Figure~\ref{fig:mgal_P_I} shows the placement of the \textit{P} and \textit{I} populations in the Mg-Al anti-correlation determined in this study. Blue indicates the \textit{P} population and purple indicates the \textit{I} population stars. The stars which lie within the uncertainties of the threshold are in orange as these stars could be attributed to either population. 

Figure~\ref{fig:mgal_P_I}h reproduces the analysis presented in Figure 3 of \citet{carretta_2014}, showing the distribution of the \textit{P} and \textit{I} populations in NGC~2808 in the Mg-Al abundance plane. The threshold between the two populations found in this work appears to align with the \citet{carretta_2014} division ([Al/Fe]$\sim0.7$). While the \textit{E} population may be present near [Al/Fe]$>$1 and [Mg/Fe]$<$0 in Figure~\ref{fig:mgal_P_I}h, this work focuses on the two primary populations found with GMM. Thus for the dual population GCs in this study, we defined only the two populations of \textit{P} and \textit{I}. 

\subsection{Metallicity in the Primordial and Intermediate Populations}
\label{sec:pop_feh}

\begin{figure*}
    \centering
    \includegraphics[width=17cm]{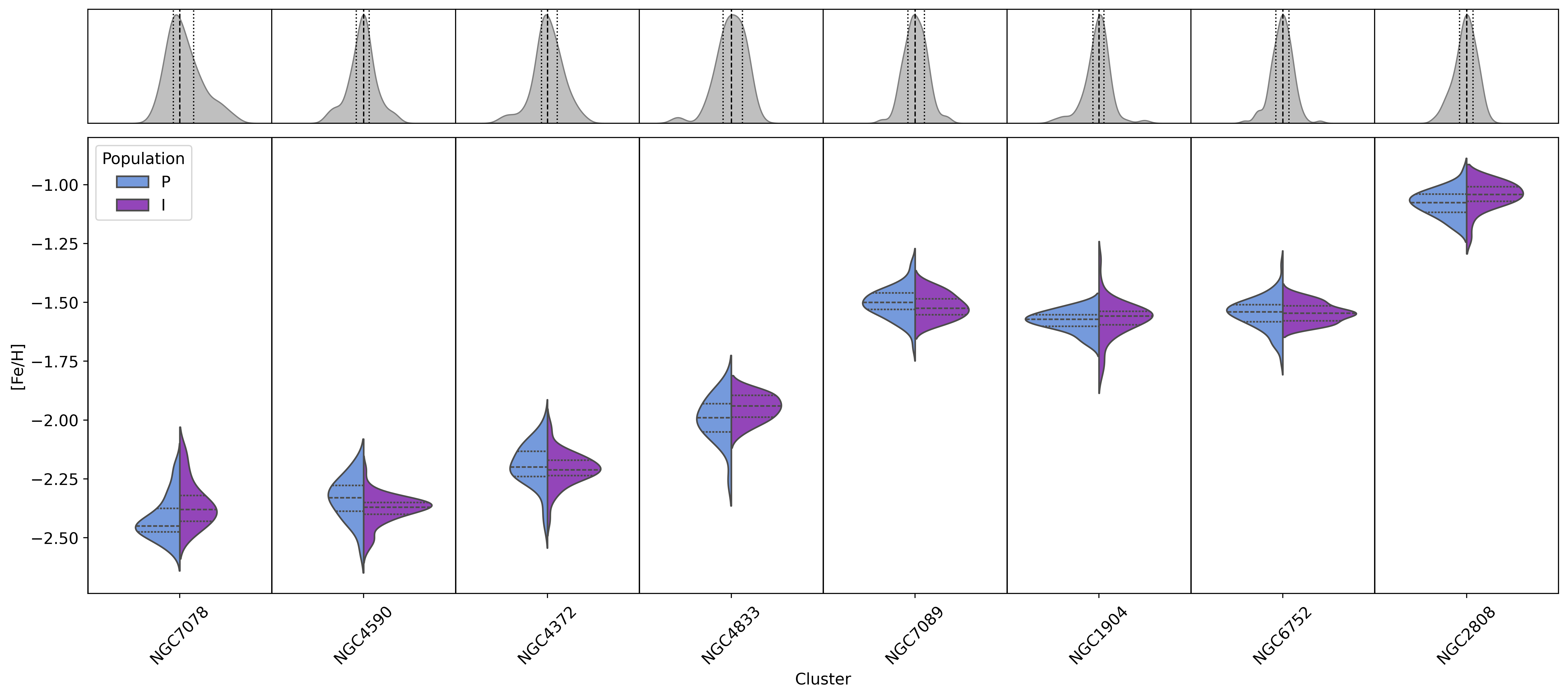}
    \caption{Top panel: Histograms for each cluster's [Fe/H] values for stars with Mg and Al measurements from this work. Lower panel: Violin plot of the \textit{P} (blue) and \textit{I} (purple) population's metallicity distribution. Dashed lines within both plots indicate the median values, dotted lines indicate the upper and lower quartile boundaries.}
    \label{fig:feh_violin}
\end{figure*}

The presence of a metallicity spread within the first generation (\textit{P} population) of GCs is a developing topic in recent studies such as \citet{cabrera2019_2808, lardo2018_spread, lardo2022_spread, bailin2022, legnardi2022_spread, latour2025_census, carretta2025_spread, schiappacasse2026}. Figure \ref{fig:feh_violin} displays the spread in [Fe/H] for both the \textit{P} (blue) and \textit{I} (purple) samples within the dual population clusters in this study. The upper panel shows the combined distribution of [Fe/H] for the stars of both populations in each cluster. Additionally, Table \ref{tab:pop_feh_spreads} displays the median [Fe/H], interquartile range (IQR), standard deviation ($\sigma$), number of stars (N$_{*}$) and [Fe/H] typical error for both populations. 

To assess any separation between the populations in [Fe/H], the difference between the median metallicity in \textit{P} and \textit{I} was investigated. This difference in [Fe/H] (Med $\Delta$), the corresponding t-statistic and p-value are also shown in Table \ref{tab:pop_feh_spreads}. While NGC~7078 has the largest, statistically significant difference, as seen by the low p-value combined with a t-statistic relatively far from 0, this value is less than the typical error. Therefore, it cannot be confidently said that there is any separation in [Fe/H] between \textit{P} and \textit{I}.

Metallicity spreads within each population implies even more chemically complex histories and the possibility of further sub-populations. In the literature, the spread can be reported as a range (e.g., max-min, IQR), or a standard deviation. As noted above, we report both IQR and $\sigma$ in Table \ref{tab:pop_feh_spreads}. Here we discuss the IQR as shown in Figure \ref{fig:feh_violin}, as the distributions are generally skewed, while the $\sigma$ assumes a Gaussian distribution, and an IQR describes the range of metallicity of the central 50$\%$ of the sample. The IQR for both \textit{P} and \textit{I} populations in NGC~7078 and NGC~4833 are all larger than the typical errors (Err). This is also true for the \textit{P} populations in NGC~4590, NGC~2808 and NGC~4372. When the IQR is larger than the error, this is evidence for an astrophysical source to explain the metallicity spread, especially for NGC~4833.

While literature is sparse on the I population \citep{carretta2025_spread, schiappacasse2026}, studies of first population (\textit{P}) stars in GCs have found differing metallicity spreads, where photometric studies often find larger spreads than spectroscopic studies, as discussed in \citet{carretta2025_spread}. For NGC~2808 as an example, \citet{legnardi2022_spread} use photometry and reported a spread in metallicity of 0.11$\pm$0.01. Spectroscopic studies of NGC~2808 \citet{cabrera2019_2808, lardo2023_spread2808} report a spread of $\sim$0.25$\pm0.06$ (as a range from minimum to maximum). Using the data from table 1 of \citet{cabrera2019_2808}, the spread as an IQR is 0.11, and the $\sigma$ is 0.08, which is in good agreement with our values. Also, \citet{schiappacasse2026} reports $\sigma=$0.03$\pm$0.01. All of these spectroscopic studies report values for the spread on the order of the metallicity errors, in particular the $\sigma$. In all of these cases for NGC~2808, the spectroscopic spreads are smaller than or equal to those reported by photometry, in agreement with \citet{carretta2025_spread}. While the metallicity spread from photometry may imply a more complex star formation history in the first population, the spectroscopic studies do not provide convincing evidence.

We find no obvious correlation between the metallicity spread of either population and global cluster parameters such as [Fe/H] or present-day mass. Further discussion of global metallicity trends for all clusters is presented in Section \ref{sec:global_feh_s}.

\begin{table*}
\centering
\caption{Median (Med), interquartile range (IQR), standard deviation ($\sigma$), number of stars (N$_{*}$) and typical error (Err) for [Fe/H] in the \textit{P} and \textit{I} populations. The difference between the medians of \textit{P} and \textit{I} are also shown in the third last column and order the clusters by largest to smallest. T-statistics and p-values are also given in the last two columns.}
\begin{tabular}{|l|c|c|c|c|c|c|c|c|c|c|c|c|c|}
\hline
 & \multicolumn{5}{c|}{P [Fe/H]} & \multicolumn{5}{c|}{I [Fe/H]} & \multicolumn{3}{c|}{P $\&$ I [Fe/H]}\\ \cline{1-14}

Cluster & Med & IQR & $\sigma$ & N$_{*}$ & Err & Med & IQR & $\sigma$ & N$_{*}$ & Err & Med $\Delta$ & t-statistic & p-value \\
\hline\hline

NGC7078 & -2.45 & 0.10 & 0.09 & 55 & 0.08 & -2.38 & 0.11 & 0.09 & 27 & 0.08 &  0.07 & -2.81 & 0.01 \\\hline
NGC4833 & -1.99 & 0.12 & 0.09 & 21 & 0.06 & -1.94 & 0.09 & 0.05 & 22 & 0.07 & 0.05 & -1.94 & 0.06 \\\hline
NGC4590 & -2.33 & 0.11 & 0.09 & 20 & 0.07 & -2.37 & 0.05 & 0.06 & 48 & 0.07 & 0.04 & 2.06 & 0.05 \\\hline
NGC2808 & -1.08 & 0.08 & 0.06 & 39 & 0.07 & -1.04 & 0.06 & 0.06 & 24 & 0.07 & 0.03 & -2.02 & 0.05 \\\hline
NGC7089 & -1.50 & 0.07 & 0.07 & 57 & 0.07 & -1.52 & 0.07 & 0.05 & 24 & 0.07 & 0.02 & 0.93 & 0.36 \\\hline
NGC1904 & -1.57 & 0.05 & 0.04 & 24 & 0.07 & -1.56 & 0.06 & 0.08 & 46 & 0.07 & 0.01 & -0.53 & 0.60 \\\hline
NGC4372 & -2.20 & 0.11 & 0.09 & 44 & 0.07 & -2.21 & 0.07 & 0.07 & 42 & 0.07 & 0.01 & 0.90 & 0.37 \\\hline
NGC6752 & -1.54 & 0.07 & 0.07 & 68 & 0.07 & -1.55 & 0.06 & 0.04 & 48 & 0.07 & 0.01 & -0.35 & 0.73 \\ \hline

\end{tabular}
\label{tab:pop_feh_spreads}
\end{table*}

\section{Global Cluster Characteristics}
\label{sec:global_char}

\subsection{Mg-Al Anti-Correlation Spread}
\label{sec:mgal_spread}

\begin{figure}
    \centering
    \includegraphics[width=1\linewidth]{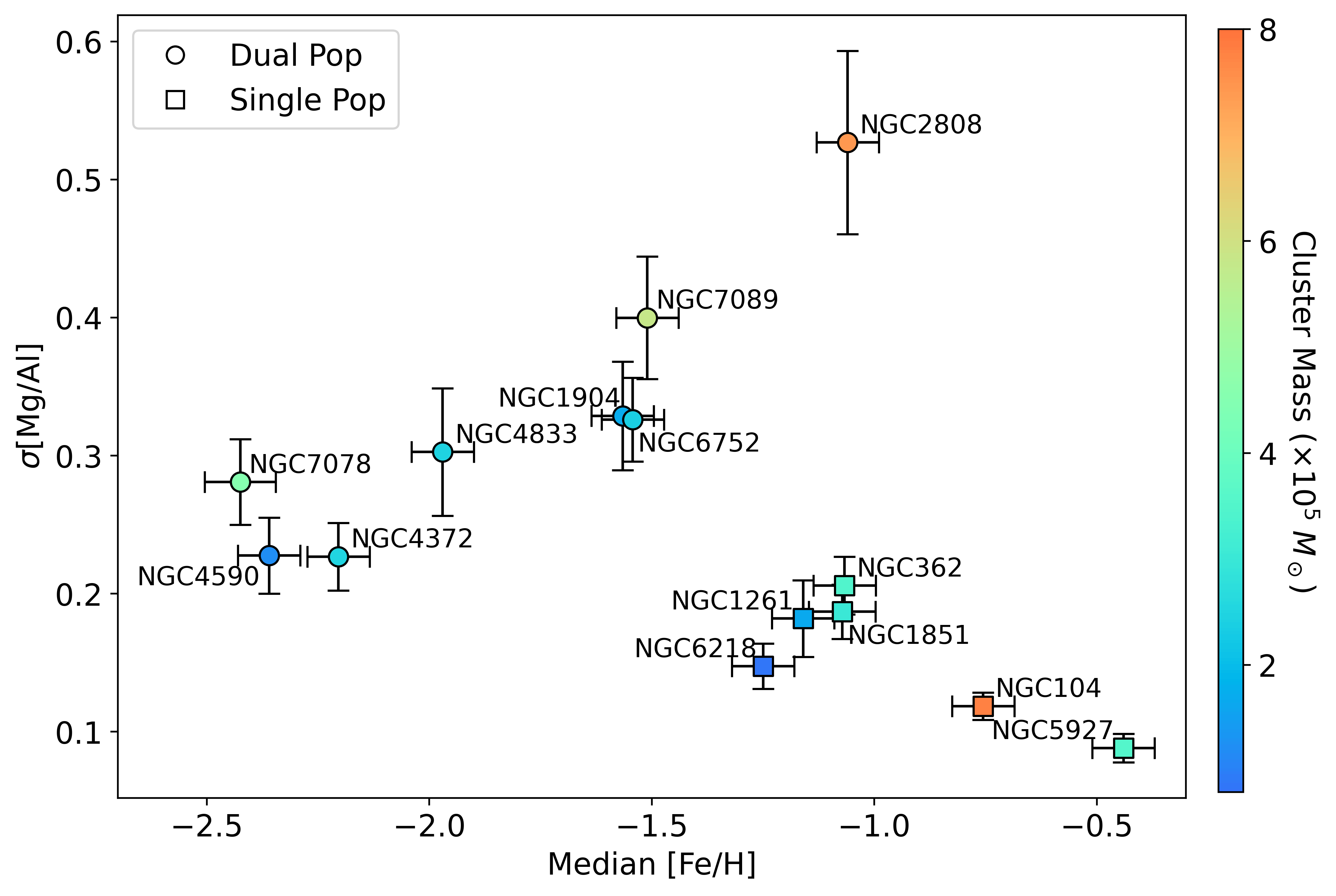}
    \caption{Mg-Al anti-correlation spread ($\sigma$[Mg/Al]) against median cluster metallicity [Fe/H] for all clusters. Dual population clusters are shown as circles, single population as squares. The colour map shows the range of present day cluster mass from \citet{mass_cat}.}
    \label{fig:mgal_cluster_params}
\end{figure}

The presence of the Mg–Al anti-correlation, and its dependence on global cluster parameters, has been explored in previous work \citep{NaOVIII, NaOVII, meszaros2015, cabrera2016, Pan_mgal, meszaros2020}. These studies show that the Mg–Al anti-correlation is predominantly found in metal poor and/or massive GCs. Figure \ref{fig:mgal_cluster_params} illustrates how the [Mg/Al] spread ([Mg/Al]$\sigma$) relates to cluster metallicity and mass for the clusters in this study. Clusters shown in red contain dual populations, while those in blue show only a single population; darker colours in each of red and blue correspond to higher masses as per the colour bars.

The dual-population clusters, except for NGC 2808, are also the most metal-poor clusters in the sample. This is only true when using the Mg-Al anti-correlation to define populations as the Na-O anti-correlation is present across all the GES clusters, regardless of metallicity. Therefore, all clusters will have two populations defined from the Na-O anti-correlation. The transition between clusters that exhibit the Mg–Al anti-correlation and those that do not occurs approximately between NGC~7089 and NGC~6218, within the metallicity range $-1.52 <$ [Fe/H] $< -1.37$. This suggests that the mechanism responsible for producing the Mg–Al anti-correlation activates typically below this metallicity. 

NGC~2808 is an exception: despite having a metallicity within the range of the single-population clusters, its high mass may provide an environment which allows the Mg–Al anti-correlation to develop. NGC~7078 also deviates from the typical trend of the dual-population clusters. Although it is the most metal-poor cluster in the sample, its larger [Mg/Al] spread is consistent with its high mass relative to other clusters of comparable metallicity. Thus, in both the cases of NGC~2808 and NGC~7078, a sufficiently large cluster mass appears capable of sustaining the Mg–Al anti-correlation or large spread in [Mg/Al], even when metallicity alone would predict a weaker signal.

In the single-population clusters, the [Mg/Al] spread decreases steadily with increasing metallicity, and this behaviour is independent of mass. NGC~104 highlights this point: despite its relatively large mass, it shows only a small [Mg/Al] spread. If the `mass–[Mg/Al] spread' relation seen in the metal-poor clusters continued into the metal-rich regime, NGC~104 would exhibit a significantly larger spread. Its position instead suggests that, above a certain metallicity, mass no longer plays a dominant role. This trend is consistent with \citet{meszaros2020} who suggest the correlation between the Al enrichment and cluster mass becomes stronger for GCs with [Fe/H]$<-1.3$.

Together, these results imply that the Mg–Al anti-correlation is most efficiently produced in low-metallicity environments, with cluster mass becoming influential only below this metallicity threshold. At higher metallicities, the mechanism driving Al enrichment either weakens or fails to operate, consistent with the conclusions of \citet{meszaros2020}. The behaviour of the clusters in the transition region, around [Fe/H]$\sim -$1.4, suggests a change in the balance of these effects, which is considered further in the context of Mg and Al nucleosynthesis in Section \ref{sec:discussion}.

\subsection{Variations in Metallicity and $s$-Process Elements}
\label{sec:global_feh_s}

The Mg-Al anti-correlation is not the only anomalous property of GCs that can indicate the presence of multiple populations, and distinguish these clusters from simple stellar populations. Several clusters also show variations in metallicity and neutron capture elements, particularly those synthesised through the slow neutron capture process ($s$-process) \citep{Fe_s_marino}. During neutron capture, light elements act as seed nuclei which can capture neutrons to produce heavier elements at either slow or rapid time scales, depending on the neutron density environment. Of particular interest are elements associated with the $s$-process, such as Y and Ba, since both can be produced in AGB stars. These stars are also thought to contribute to the Mg and Al enrichment observed in GCs \citep{ges_s_process, Kobayashi_2020, s_beyond}.

\begin{figure*}
    \centering
    \includegraphics[width=17cm]{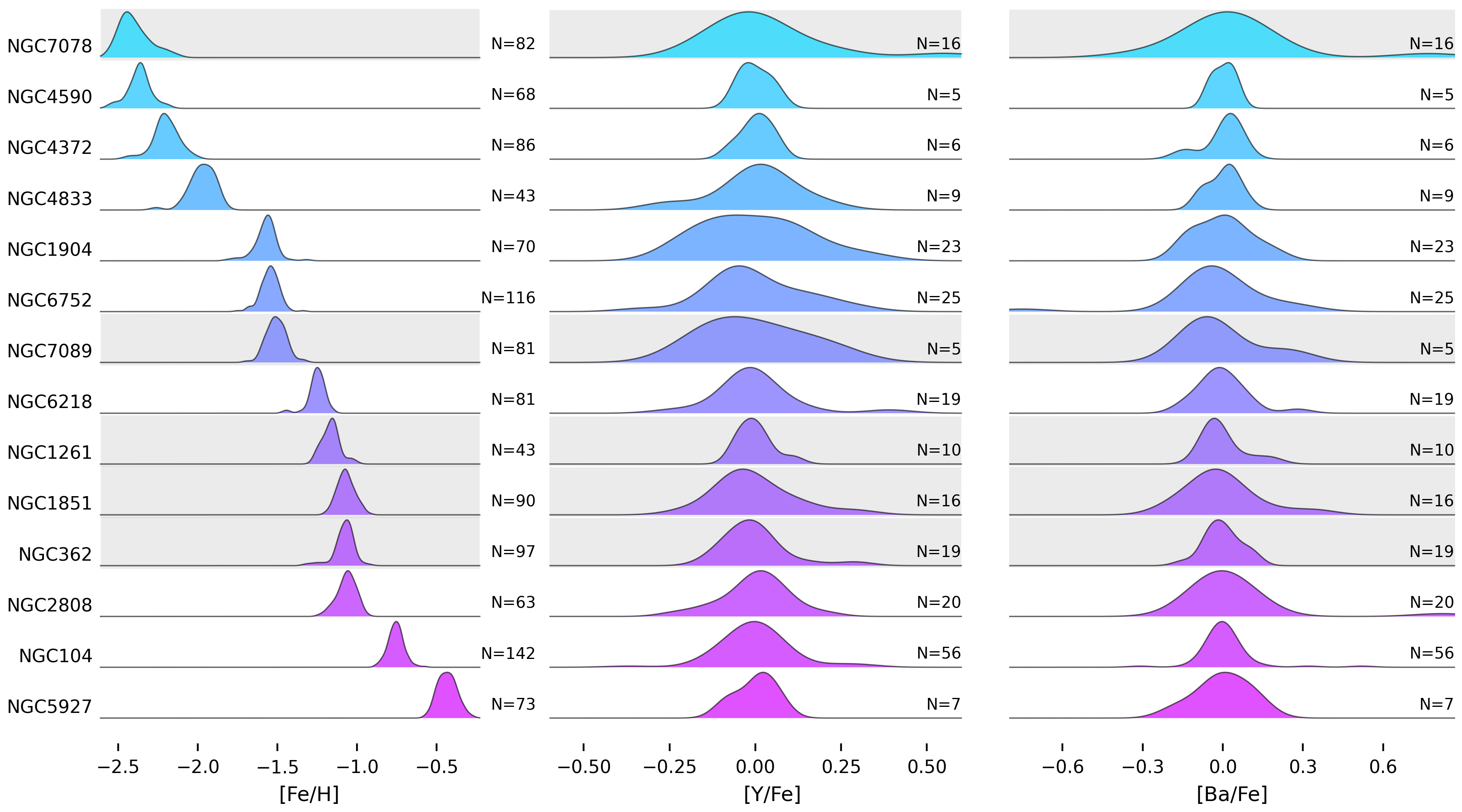}
    \caption{Distribution of [Fe/H], [Y/Fe] and [Ba/Fe] for each cluster, increasing in median metallicity from top to bottom (blue to pink). [Fe/H] are the adjusted GES values from this work, while [Y/Fe] and [Ba/Fe] are the corrected values from \citet{jose2025}. The number of stars (N) is noted to the right of each distribution. Clusters highlighted in grey are Type~\textsc{ii}}
    \label{fig:global_spreads}
\end{figure*}

To explore this broader chemical complexity, we examine metallicity distributions using our corrected [Fe/H] values and incorporate the microturbulence corrected $s$-process abundances (Ba, Y) from \citet{jose2025}. \citet{jose2025} utilise the UVES sample from GES to explore neutron capture elements with a correction for microturbulence applied. Figure \ref{fig:global_spreads} shows the distribution of [Fe/H], [Ba/Fe] and [Y/Fe] for all clusters, arranged from most metal poor to most metal rich.

Clusters exhibiting clear variations in either metallicity and/or s-process abundances have been classified as Type~\textsc{ii} by \citep{hubble_UV}. While not all Type~\textsc{ii} clusters are chemically anomalous in every element, all clusters known to host heavy element variations show Type \textsc{ii} behaviour according to their photometric chromosome maps \citep{marino2018_6934}. In our sample NGC~362, NGC~1261, NGC~1851, NGC~7089 and NGC~7078 are designated Type \textsc{ii} \citep{hubble_UV, nardiello2018_M15}, and are highlighted in grey in Figure~\ref{fig:global_spreads}. The remaining clusters are classified as Type \textsc{i}, with the exception of NGC~1904 and NGC~4372, which are yet to be assigned a type. 

The most pronounced Type~\textsc{ii} behaviour can be seen for NGC~7078, with large spreads in metallicity and both s-process elements. The broad distribution in [Y/Fe] and [Ba/Fe] are consistent with NGC~7078's well known variations in neutron capture abundances from previous spectroscopic studies \citep{sobeck_m15, worley2013ba, garcia_m15}. The tail toward higher metallicities (-2.3$\le$[Fe/H]$\le$-2.1) has not been widely reported, although subtle high-metallicity extensions are present in \citet{sneden1997_m15, sneden2000_m15, worley2013ba}, and contributes to the large total metallicity spread.

NGC~7089 also shows Type~\textsc{ii} characteristics, although this is more evident in the broader [Y/Fe] distribution compared to [Ba/Fe]. A bi-modality can be seen for both $s$-process elements and a small metal rich component is visible near [Fe/H]$\sim -$1.2. Given the limited number of stars with Y and Ba measurements, these distributions likely underestimate the true, larger variation, which has been seen in literature \citep{lardo2013_m2, yong2014_m2}.

For NGC~1851, we recover the expected large spread for both $s$-process elements \citep{yonggru2007_1851, carretta_1851, carretta2011_1851}. While there is a small, secondary peak toward [Fe/H]$>-$1, we do not find the known significant spread in metallicity \citep{carretta_1851, ges_1851, schiappacasse2026}. We also do not find the previously reported metallicity spread for NGC~1261 \citep{marino2021_1261, munoz2021_1261}. Both clusters are considered to be Type~\textsc{ii} and while this aligns with the $s$-process distribution of NGC~1851, NGC~1261 does not show any Type~\textsc{ii} characteristics in this dataset. 

The majority of Type~\textsc{i} clusters behave as expected. NGC~6218, NGC~104 and NGC~5927 all show narrow to moderate spreads in [Fe/H], [Y/Fe] and [Ba/Fe], consistent with their classification. Slightly broader than expected [Y/Fe] distributions in NGC~6218 and NGC~104 are found but remain within the range observed in other Type~\textsc{i} clusters. While a bi-modality can be seen in NGC~5927, this is likely due to the limited sample of 6 stars.

NGC~4590 and NGC~4372 also show their known Type~\textsc{i} behaviour in Figure~\ref{fig:global_spreads}, with narrow spreads across [Fe/H], [Y/Fe] and [Ba/Fe]. There does appear to be a bi-modality in the $s$-process distributions for both clusters, indicating possible $s$-rich and $s$-poor populations. This is also true for NGC~4833 which exhibits a relatively large [Fe/H] spread, that has not been reported in previous studies \citep{carretta_4833, roederer_4833}. Given the low number of stars with $s$-process abundances, the presence of two clear $s$-process populations cannot be accurately determined. 

While NGC~4590 and NGC~4372 show small spreads in the $s$-process elements, their metallicity distributions are wider than expected, similar to NGC~4833. NGC~4590 shows a central peak with tails toward both the higher and lower [Fe/H] values, increasing the overall spread. NGC~4372 however appears to have a broad central peak that is possibly bimodal.

More significant $s$-process spreads appear in NGC~1904 and NGC~6752 both of which are Type~\textsc{i} clusters. For NGC~6752 in particular, the $s$-process distributions are unusually broad, similarly for NGC~1904, showing a large spread in [Y/Fe] and [Ba/Fe]. \cite{jose6752} also investigated $s$-process elements in NGC~6752 but did not find a significant spread, however their sample was smaller than the one of \cite{jose2025} that we use in this work. While further investigation is required to confirm any $s-$process spreads in Type~\textsc{i} clusters, this may suggest that $s-$process spreads are not exclusive to Type~\textsc{ii} clusters.

Examined together, the metallicity and $s$-process abundance distributions reveal that heavy element anomalies are not restricted to Type~\textsc{ii} clusters. These results reinforce the view that multiple, chemically distinct populations are a fundamental property of GCs. The different behaviours of [Fe/H], [Y/Fe] and [Ba/Fe] highlight the need to consider elements produced in a range of nucleosynthetic sites when tracing the formation and evolution of GCs.

\section{Discussion}
\label{sec:discussion}

Light element variations in GCs are believed to be signatures of high-temperature H-burning, particularly from the CNO, NeNa and Mg-Al cycles \citep{deep_mix1993, mult_stell_pop, whatGC, MP_clust}. At low core temperatures (T$\sim$20MK), only the CN cycle occurs. However, when the core temperature rises (T$\ge$30MK), the NO cycle can activate \citep{HBurn}. Following the CNO cycle, the NeNa and Mg-Al cycles can occur when temperatures reach T$\sim$40MK and T$\sim$70MK respectively \citep{mult_stell_pop}.

The light element variations in GCs are often attributed to the CNO, Ne-Na and Mg-Al cycles where each stage activates once higher temperatures are reached. \citet{nataf2019} highlight another possibility, where CNO and Ne-Na form one pathway while Mg-Al represents a separate process. Distinguishing between these scenarios therefore requires examining variations in Na, O, Mg and Al together.

Since the Mg-Al cycle requires temperatures higher than those reached in the interiors of the low-mass stars observed today, it cannot be responsible for the abundance variations produced during their current evolutionary phases. Therefore, the Mg and Al abundance trends observed in GCs are not evidence of nucleosynthesis occurring during normal low-mass stellar evolution \citep{mult_stell_pop}.

Three primary stellar types are known to have interiors sufficiently extreme to activate the Mg-Al cycle. These are massive stars with M$\ge$15M$_\odot$ \citep{superbubble}, very massive stars with M$\sim10^{4}$M$_\odot$ \citep{Denissenkov_Hartwick_2013} and intermediate mass (M$\sim$3$-$8M$_\odot$) AGB stars experiencing Hot Bottom Burning \citep{single_model}. These higher mass stars were likely present in the first generation of GC stars, which have already evolved and died to pollute the second generation of stars we see today \citep{whatGC}.

One of the leading scenarios proposed to explain the observed abundance variations in GCs involves enrichment by AGB stars. As the Mg-Al cycles progresses in the H-burning shell of AGB, Mg will become depleted as it is fused into Al, therefore Al is enriched, producing the observed Al and Mg abundance spread \citep{karakas2003}. While this may explain the Mg-Al anti-correlation, all current AGB models produce a Na-O correlation rather than the commonly observed anti-correlation \citep{mult_stell_pop}. 

To account for the missing O poor stars, dilution theories, such as the accretion of pristine material to be mixed with the AGB ejecta and create O-poor stars, have been considered \citep{dercole2016}. There are many complexities with these models such as the origin and timing of the dilution material being difficult to explain \citep{mult_stell_pop}. Some form of dilution is required for most polluters, not just AGB, such as fast-rotating massive stars and very massive stars, indicating GC formation may be complicated by other mechanisms we are yet to understand \citep{whatGC}.

AGB stars are also key $s-$process sites as they produce both the light and heavy $s-$process elements. The mass of the AGB stars within GCs can be constrained using $s-$process elements as the light $s$ (e.g. Y) are generated within intermediate mass AGB, compared to the heavy $s$ (e.g. Ba) which are generated within low mass AGB \citep{Kobayashi_2020}. The amount of $s-$process elements measured in a star also can indicate the metallicity of the progenitor AGB. Most clusters we found to show $s-$process variations were based on the Y abundances, hence there is more contribution of light $s$-process elements which indicates the influence of intermediate mass AGB stars. These clusters include NGC~7078, NGC~4833, NGC~1904, NGC~6752, NGC~6218 and NGC~104.

Although the metal rich clusters mentioned in Section \ref{sec:stat_method} were excluded from the population investigation, they are known to contain multiple populations based on the Na-O anti-correlation, which is present in all GES clusters. Specifically, populations based on the Na-O anti-correlation for NGC~104, NGC~362, NGC~1851, NGC~1904, NGC~2808, NGC~4590, NGC~6752, NGC~6218 and NGC~7078 have been identified in \citet{NaOVIII, NaOVII, carretta2011_1851, carretta2013_362}. \citet{Pan_mgal} presents both Na-O and Mg-Al anti-correlations for NGC~104, NGC~362, NGC~1851, NGC~1904, NGC~2808, NGC~4833, NGC~6752, NGC~7089 and the first report of these anti-correlations in NGC~5927 from GES iDR4. The only report of the Na-O anti-correlation for NGC~1261 can be seen in \citet{marino2021_1261}. 

The GES dataset is particularly valuable for probing these questions, as it provides a large homogeneous sample across many clusters. However, while it contains extensive Mg and Al coverage, Na and especially O are missing due to the choice of GIRAFFE setups in which GES observed, as mentioned in Section \ref{sec:final_abunds}. This limits our ability to trace both anti-correlations simultaneously and the connection between the nucleosynthetic cycles cannot be fully tested. Measuring Na and O abundances for all clusters would enable us to investigate the populations in metal-rich globular clusters in relation to Mg and Al. 

The presence of the Mg-Al anti-correlation in some clusters and not others shows that further investigation is needed into the conditions that activate the Mg-Al cycle. Based on our analysis, mass and metallicity appear to be key drivers of these conditions, in agreement with conclusions in the literature. Expanding the dataset to include Na and O, would allow the light element cycles to be linked more securely and help to clarify whether light element variations in GCs arise from a continuous chain of high-temperature H-burning or from distinct nucleosynthetic pathways.

\section{Conclusion}
\label{sec:conc}

In this work, we have extended the GES GC sample by deriving additional Mg and Al abundances, enabling a more complete exploration of the Mg-Al anti-correlation across 14 clusters. We applied a GMM to the [Mg/Al] distribution to divide clusters where a large variation in the Mg-Al distribution is present, into two primary stellar populations. 

While NGC~7078, shows a differences in metallicity between the two populations, this is less than the typical [Fe/H] error. High precision measurements with lower uncertainties are required to confirm the possible [Fe/H] separation and population spreads seen in this work. If confirmed, this could suggest that metallicity variations within populations may be more common than previously thought, though their origin remains unclear.

We confirm that the presence and spread of the Mg-Al anti-correlation depends strongly on cluster mass and metallicity. Clusters that are both massive and metal-poor are more likely to show a pronounced anti-correlation. In contrast, clusters more metal-rich than [Fe/H]$\sim -$1.4 generally lack this chemical feature, indicating a transition region where the mechanism driving Al enrichment and Mg depletion appears to deactivate. The Mg-Al anti-correlation morphology across the metallicity range shows the term 'anti-correlation' is currently very broad, and does not cover everything we observe. 

We also see in the dual population clusters, the [Mg/Al]$\sigma$ increases with metallicity, while in the single population clusters (which reside in a higher metallicity regime) it decreases with metallicity. Comparing this behaviour to the Na-O anti-correlation will help to distinguish between different CNO-cycle stages as Na-O is observed across GCs, while Mg-Al activates only for a specific mass and metallicity range.

Beyond the light elements, several clusters show spreads in global metallicity (e.g., NGC~4833, NGC~4372 and NGC~7078) and $s$-process elements (e.g., NGC~7078, NGC~1904, NGC~6752). While some of these agree with previous findings, others do not align with current classifications of Type~\textsc{i} and Type~\textsc{ii} clusters. Expanding the $s-$process sample with more measurements and a greater range of elements would allow further investigation into the validity of the detected spreads.

Overall, this work demonstrates the value of homogeneous, large scale datasets such as GES for investigating the chemical complexity of GCs. Combining both Mg-Al and Na-O anti-correlations in future work, alongside expanded $s$-process measurements, will provide critical insights into the nucleosynthetic origins of these abundance variations and multiple populations. These results reinforce that GCs are far from simple systems with straightforward formation histories. Furthering our understanding of the nature of these objects, and thus the impact on galactic chemical evolution, is essential.

\begin{acknowledgements}
      We thank the anonymous referee for their valuable comments. We also thank P. Cottrell and Q. Aicken-Davies for their helpful comments and suggestions. Based on data products from observations made with ESO Telescopes at the La Silla Paranal Observatory under programmes 188.B-3002, 193.B-0936, and 197.B-1074. This work has made use of Python packages SciPy \citep{SciPy}, NumPy \citep{numpy}, SciKitLearn \citep{scikit-learn}, Matplotlib \citep{matplotlib} and AstroPy \citep{astropy}. 
      H.S.W thanks Claudia Aguilera Gomez for the invitation to present this work at Pontificia Universidad Católica de Chile and the funding for a research visit. J.S.U. thanks INAF for the support through the Mini-Grant (1.05.24.07.02). P.J thanks for the support from FONDECYT REGULAR 1231057. 
      J.S.U. and L.M. acknowledge support from INAF through the Large Grants EPOCH, funding for the WEAVE project, the Mini-Grants Checs (1.05.23.04.02), and financial support under the National Recovery and Resilience Plan (PNRR), Mission 4, Component 2, Investment 1.1, Call for tender No. 104 published on 2 February 2022 by the Italian Ministry of University and Research (MUR), funded by the European Union – NextGenerationEU, through the Project ‘Cosmic POT’ (Grant Assignment Decree No. 2022X4TM3H, MUR).
\end{acknowledgements}

\bibliographystyle{aa}
\bibliography{Ref}

\appendix

\section{Metallicity Corrections in GIRAFFE}
\label{apdx:feh_corr}

The most significant metallicity trends were seen for the GIRAFFE stars. As part of the GES homogenisation, the GIRAFFE results were scaled to UVES using a reference set of stars in common \citep{Worley_2024}. \citet{Worley_2024} reported that for the GCs sample, there were not many stars in common between GIRAFFE and UVES and so the metal poor end was poorly sampled. As the GCs were not homogenised separately, it is unsurprising that some trends with parameters exist for GCs in the GIRAFFE sample.

To address these trends, a linear fit was therefore applied to the stars in the cluster and subtracted based on the parameter with the strongest trend. This was required for 8 clusters, where each had an individual fit and correction applied. The clusters with $T_{\text{eff}}$ corrections are NGC~4372, NGC~1851, NGC~2808 and $\log g$  corrections are NGC~1904, NGC~6752, NGC~6218, NGC~1261, NGC~104. While the trends were defined as statistically significant with p-values $<$ 0.05, the median of the corrections in metallicity for the stars across all clusters was less than 0.025 dex. This is less than the median metallicity error per star of 0.07 dex. Only two stars required corrections larger than their metallicity error, with the largest correction 0.075 dex, while the error is 0.06 dex. These corrections are therefore expected to have a comparably small affect on the abundance derivation.

The GCs were not a primary science objective in GES and as the stars are mostly metal-poor, bright giants, they cover the parameter range for which the GES analysis is not as robust \citep{Worley_2024}. We apply a new correction for GIRAFFE here, inspired by the GES analysis, to remain on the GES scale and keep homogeneity across the clusters.

\section{Abundance Dataset}\label{apdx:A}

\begin{table*}
\centering
\caption{First five object IDs and their T$_{\text{eff}}$, $\log g$, [Fe/H], v$_{\text{t}}$, correction in [Fe/H] and abundances with respective errors of stars used in this work. This is available in full through CDS.}
\label{tab:example_dataset}
    \begin{tabular}{|l|l|l|p{1.3cm}|p{1.2cm}|p{0.7cm}|p{0.8cm}|p{0.8cm}|p{0.8cm}|p{1cm}|p{1cm}|}
    \hline
    Object & RA & Dec & Cluster & T$_{\text{eff}}$ (K) & $\delta$T$_{\text{eff}}$ (K) & $\log g$ (dex) & $\delta$$\log g$ (dex) & [Fe/H] (dex) &  $\delta$[Fe/H] (dex) & $\Delta$[Fe/H] (dex) \\\hline
    00214385-7207213 & 5.432708 & -72.122583 & NGC104 & 5017 & 62 & 2.94 & 0.14 & -0.74 & 0.08 & 0.01 \\\hline
    00221154-7200452 & 5.547833 & -72.012556 & NGC104 & 5125 & 66 & 2.89 & 0.14 & -0.75 & 0.07 & 0.01\\\hline
    00222583-7207070 & 5.607583 & -72.118667 & NGC104 & 4629 & 63 & 2.11 & 0.14 & -0.78 & 0.08 & 0.01\\\hline
    00223022-7212296 & 5.625917 & -72.208222 & NGC104 & 4984 & 62 & 2.82 & 0.14 & -0.75 & 0.17 & 0.01 \\\hline
    00223198-7206489 & 5.633208 & -72.113583 & NGC104 & 4472 & 16 & 1.64 & 0.10 & -0.75 & 0.07 & 0 \\ \hline
    \end{tabular}

    \vspace{0.5cm}

    \begin{tabular}{|l|p{1.3cm}|p{1cm}|p{1.2cm}|p{1cm}|p{1.1cm}|l|}
    \hline
    Object & v$_t$ (kms$^{-1}$) & [Mg/Fe] (dex) & $\delta$[Mg/Fe] (dex) & [Al/Fe] (dex) & $\delta$[Al/Fe] (dex) & Instrument \\\hline
    00214385-7207213 & 1.27 & 0.37 & 0.04 & 0.53 & 0.05 & GIRAFFE  \\\hline
    00221154-7200452 & 1.36 & 0.39 & 0.03 & 0.54 & 0.06 & GIRAFFE \\\hline
    00222583-7207070 & 1.51 & 0.33 & 0.02 & 0.30 & 0.06 & GIRAFFE\\\hline
    00223022-7212296 & 1.32 & 0.31 & 0.03 & 0.42 & 0.16 & GIRAFFE \\\hline
    00223198-7206489 & 1.67 & 0.53 & 0.02 & 0.45 & 0.13 & UVES \\ \hline
    
    \end{tabular}
\end{table*}

\end{document}